\documentclass[amsmath,amssymb,showpacs,draft,preprint,floatfix]{revtex4}
\usepackage{graphicx}
\usepackage{latexsym}
\begin{document}

\title{\textbf{Quantum bits and superpositions displaced Fock states of the cavity field}}
\author{L.M. Ar\'evalo Aguilar$^1$ and H. Moya-Cessa$^2$}
\date{\today}\address{$^1$Centro de Investigaciones en Optica,
A.C., Prolongaci\'on de Constituci\'on No. 607,Apartado Postal No.
507, Fracc. Reserva Loma Bonita, 20200 Aguascalientes, Ags.\\
$^2$Instituto Nacional de Astrof\'{\i}sica, Optica y
Electr\'onica, Apdo. Postal 51 y 216, 72000 Puebla, Pue.}

\begin{abstract}
We study the effects of counter rotating terms in the interaction
of quantized light with a two-level atom, by using the method of
small rotations.  We give an expression for the wave function of
the composed system atom plus field and point out one initial wave
function that generates a quantum bit of the electromagnetic field
with arbitrary amplitudes.

\ \ \ \ \ \ \

{\bf Resumen}
\ \ \

Estudiamos los efectos de los t\'erminos contra-rotantes en la interacci\'on entre un
campo elctromagn\'etico cuantizado y un \'atomo de dos nivels, usando el m\'etodo de
peque\~nas rotaciones. Obtenemos una expresi\'on para la funci\'on de onda del sistema
compuesto \'atomo-campo y elegimos un estado inicial de la funci\'on de onda que genera un
bit cu\'antico del campo electromagn\'etico.
\end{abstract}

\maketitle
\section{introduction}
The Jaynes-Cummings model (JCM) \cite{Knight} is a very simplified
version of a much more complex problem, the interaction between
electromagnetic radiation and atoms. It models this interaction
using the rotating wave approximation (RWA) that allows it to be
fully solvable. Its simplicity allows physicists to apply the
fundamental laws of quantum electrodynamics to it and still be
able to solve it analytically. In the early days of its existence,
the JCM was regarded as a theoretical curiosity because of the
inherent difficulties in its experimental realization. Over the
past few years, however, there have been a number of experiments
\cite{exp} that can be modeled by the Jaynes-Cummings Hamiltonian
or generalizations of it \cite{Dav}. Therefore, the JCM has been a
subject of great interest because it enables one to study, in a
realistic way, not only the coherent properties of the quantized
field, but also its influence on atoms. Collapses and revivals
\cite{san}, squeezing \cite{kuk}, generation of Sch\"odinger cat
states \cite{gea}, etc. have been predicted with this model, and
recent developments in Cavity QED techniques have made it possible
to observe those phenomena \cite{exp}. Moreover, there has been a
{\it revival} of the JCM because its Hamiltonian can be used to
model some other systems, such as the interaction of a trapped ion
with a laser field. On this topic, multiphonon and anti JCM may be
produced \cite{win1,kis,mato,moyac,winl}, giving rise to a variety
of phenomena thought technologically difficult to realize in
cavity QED. Recent advances in quantum information processing have
given importance to quantum state engineering, as one needs to
produce and control quantum bits or qubits (superpositions of the
ground and first excited level) of the system (see for instance
\cite{moyac}). However, quantum noise, that destroys quantum
coherences very fast, can be very difficult to overcome, and ways
of protecting states have been published \cite{eke}. Here we would
like to treat the problem of a two-level atom interacting with a
quantized field but not considering the RWA, because a state
produced by the JCM (with RWA) could be thought as if it had some
noise (because a small correction with a further evolution can
mislead a desired result). Therefore, it is important to give the
most exact possible solution to the problem of interaction of a
two-level atom with a quantized field. Recently some eigenstates
for the complete Hamiltonian have been found \cite{moy3}. However,
as they do not form a complete basis, exact solutions may be found
only  for those (eigen) states (that of course may be regarded as
trapping states).

Former studies on the effects of counter rotating \cite{zah,zah1}
terms have used the path integral technique in coherent state
representation (the coherent state propagator) and have shown that
even under conditions in which the RWA is considered to be
justified, there is significant contribution to the atomic
inversion due to counter rotating terms \cite{zah1}. They obtained
results to first order for the atomic inversion \cite{zah1} and
the average number of photons \cite{zah}, however,  there is no
explicit result for the wave function. We believe that it is
important to have expressions that are easy to manipulate,
specifically of the wave function, because of the possibility of
generating non-classical states of light, which in the past has
open the field of quantum state engineering. Therefore, we
reconsider the problem in this manuscript, and show that
non-classical states, namely, qubits  and superpositions of
displaced number states \cite{cessa} of the quantized
electromagnetic field may be produced.

We study the problem from the point of view of the small rotations
method proposed recently by Klimov and S\'anchez-Soto \cite{kli},
to obtain a first order correction to the wave function. The RWA
breaks down as the atomic frequency and the field frequency are
detuned, and we consider this (detuned)  case. With the expression
for the evolved wave function, we show what initial state has to
be used in order to obtain a qubit of the quantized field with
arbitrary amplitudes. In Section II we transform the Hamiltonian
for the atom-filed interaction by applying the method of small
rotations to have an effective Hamiltonian that can be fully
solved. In section III we apply it to an initial wave function in
order to obtain qubits, and displaced superpositions of Fock
states and section IV is devoted to conclusions.
\section{Introduction}
The Hamiltoninan for the system of a single two-level atom interacting with a
single-mode quantized field in the dipole approximation is given by (we have set $\hbar=1$)
\cite{Knight}

\begin{equation}
\hat{H}=\omega \hat{n}+\frac{\omega_0}{2} \hat{\sigma}_{z}+
\lambda (\hat{\sigma}_{+} +\hat{\sigma}_{-})
(\hat{a}+\hat{a}^{\dagger }).
\label{NJCM}
\end{equation}
where $\hat{a}^{\dagger}$ and $\hat{a}$ are the creation and
annihilation operators for the field mode, respectively, obeying
$[\hat{a},\hat{a}^{\dagger}]=1$, $\hat{n} =
\hat{a}^{\dagger}\hat{a}$ and $\hat{\sigma}_+=|e\rangle\langle g|$
and $\hat{\sigma}_-=|g\rangle\langle e|$ are the raising and
lowering atomic operators, respectively, $|e\rangle$ being the
excited state
 and $|g\rangle$ the ground state of the two-level atom.
The atomic operators obey the commutation relation
$[\hat{\sigma}_+,\hat{\sigma}_-]=\hat{\sigma}_z$. $\omega$
is the field frequency, $\omega_0$ the atomic frequency and $\lambda$ is the
interaction constant.

We apply the transformation
\begin{equation}
\hat{T}=e^{-\delta (\hat{a}-\hat{a}^{\dagger})(\hat{\sigma}_{+} + \hat{
\sigma}_{-})},  \label{tra2}
\end{equation}
to equation (\ref{NJCM}) to obtain
\begin{eqnarray}
\hat{\cal{H}}  & = & \omega ( \hat{n}-\delta (\hat{a}+\hat{a}^{\dagger})
(\hat{\sigma}_{-}+\hat{\sigma}_{+})+\delta^2)  \nonumber \\
&  & +\frac{\omega_0}{2} \left( \sigma_{z}\cosh [2\delta (\hat{a}-\hat{a}^{\dagger})]
-(\sigma _{-}-\sigma_{+})\sinh [2\delta (\hat{a}-\hat{a}^{\dagger})] \right)  \nonumber \\
 & & + \lambda (\hat{a}+\hat{a}^{\dagger })(\hat{\sigma}_{-}+\hat{\sigma}_{+})-2\lambda\delta
\label{h2}
\end{eqnarray}
by considering the quantity $\delta$ much smaller than one, we can approximate
(\ref{h2}) to first order
\begin{eqnarray}
\hat{\cal{H}} & \approx & \omega \left( \hat{n}-\delta (\hat{a}+\hat{a}^{\dagger })(
\hat{\sigma}_{-}+\hat{\sigma}_{+})\right)
+\frac{\omega_0}{2} \left( \sigma_{z}+2\delta (\sigma _{+}-\sigma _{-})(\hat{a}-\hat{a}^{\dagger
})\right)  \nonumber \\
   &   & +   \lambda(\hat{a}+\hat{a}^{\dagger })(\hat{\sigma}_{-}+\hat{\sigma}_{+}),
\end{eqnarray}
where we dropped constant terms that contribute to a shift of the overall
energy.
By setting
\begin{equation}
\delta =\frac{\lambda }{\omega +\omega_0 },
\end{equation}
we finally obtain a Hamiltonian similar to the one obtained when the
RWA is applied
\begin{equation}
\hat{\cal{H}}=\omega \hat{n}+\frac{\omega_0}{2} \hat{\sigma}_{z}+\epsilon (\hat{\sigma}_{+}
\hat{a}+\hat{a}^{\dagger }\hat{\sigma}_{-}).
\label{JCM}
\end{equation}
However it should be noticed that the {\it new } interaction
constant $\epsilon$ has changed from the initial one ($\lambda$),
something that does not occur when the RWA is applied. The new
interaction constant is now
\begin{equation}
\epsilon= \lambda\frac{2\omega_0}{\omega_0+\omega}.
\end{equation}
The expression for $\delta$ is exactly equal to the expression for
the first order approximation used in \cite{zah,zah1} using path
integral approach to the problem in the $\omega=\omega_0$. Note
that both methods give first order approximations and the
expansion parameter here agrees with reference \cite{zah,zah1}. We
would like however to stress that the present method allows
visualization of the form that the evolved wave function that
allows the generation of some non-classical states.

The evolved wave function may be found now by applying the
transformed unitary evolution operator to an initial wave
function:
\begin{equation}
|\Psi (t)\rangle= \hat{T}^{\dagger} \hat{U} \hat{T} |\Psi (0) \rangle,
\label{solu}
\end{equation}
where $\hat{U}$ is given by
\begin{equation}
\hat{U}=e^{-it(\omega\hat{n}+\frac{1}{2}\omega_0\hat{\sigma}_{z})} e^{-it[\frac{\Delta}{2}\hat{
\sigma}_{z}+\epsilon( \hat{a}\hat{\sigma}_{+}+\hat{a}^{\dagger}\hat{\sigma}
_{-})]},  \label{unitario}
\end{equation}
where $\Delta=\omega_0-\omega$. Equation (\ref{unitario}) may be re-written as
\begin{equation}
\hat{U}= e^{-it(\omega\hat{n}+\frac{1}{2}\omega_0\hat{\sigma}_{z})}
\left( \frac{1}{2}[\hat{U}_{11}+\hat{U}_{22}]\hat{I} + \frac{1}{2}[
\hat{U}_{11}-\hat{U}_{22}]\hat{\sigma}_z + \hat{U}_{21}\sigma_- +
\hat{U}_{12}\sigma_+ \right),
\end{equation}
where
\begin{equation}
\hat{U}_{11}(t;\hat{n}) = \cos\hat{\Omega}_{\hat{n}+1}t-i\frac{\Delta}{2} \frac{\sin\hat{\Omega}_{
\hat{n}+1}t}{\hat{\Omega}_{\hat{n}+1}},
\end{equation}
\begin{equation}
\hat{U}_{12}(t;\hat{n}) = -i\epsilon\hat{a} \frac{\sin\hat{\Omega}_{\hat{n}}t}{\hat{\Omega
}_{\hat{n}}},
\end{equation}
\begin{equation}
\hat{U}_{21}(t;\hat{n}) = -i\epsilon\hat{a}^{\dagger} \frac{\sin\hat{\Omega}_{\hat{n}+1}t}
{\hat{\Omega}_{\hat{n}+1}},
\end{equation}
and
\begin{equation}
\hat{U}_{22}(t;\hat{n}) = \cos\hat{\Omega}_{\hat{n}}t+i\frac{\Delta}{2} \frac{\sin\hat{\Omega}_{
\hat{n}}t}{\hat{\Omega}_{\hat{n}}},
\end{equation}
with
\begin{equation}
\hat{\Omega}_{\hat{n}} =\sqrt{\frac{\Delta^2}{4}+\epsilon^2\hat{n} }.
\end{equation}

\section{qubits of the quantized field}
If the cavity field is initially prepared in the coherent state $|-\delta\rangle$,
\begin{equation}
|-\delta\rangle = \hat{D}(-\delta)|0\rangle= e^{-\frac{\delta^2}{2}}
\sum^{\infty}_{n=0}\frac{(-\delta)^n}{\sqrt{n!}}|n\rangle
\end{equation}
where $\hat{D}(-\delta)=\exp[-\delta(\hat{a}^{\dagger}-\hat{a})]$ and $|n\rangle$ is the Fock number
state of the cavity field, and the atom is prepared in a superposition of the excited and ground states,
the initial (total) wave function may be written as
\begin{equation}
|\Psi (0) \rangle= \frac{1}{\sqrt{2}}(|g\rangle + |e\rangle) |-\delta \rangle .
\label{psi}
\end{equation}
By inserting (\ref{psi}) in (\ref{solu})  we obtain the entangled state
\begin{equation}
|\Psi (t) \rangle =  \frac{1}{\sqrt{2}} (|\psi_e \rangle |e\rangle +
|\psi_g \rangle |g\rangle) ,
\label{psit}
\end{equation}
with
\begin{eqnarray}
|\psi_e \rangle  & = &
(U_{11}(t;0)e^{-i\frac{\omega_0 t}{2}} \cosh[\delta(\hat{a}-\hat{a}^{\dagger})]+
U_{22}(t;0)e^{i\frac{\omega_0 t}{2}} \sinh[\delta(\hat{a}-\hat{a}^{\dagger})])|0 \rangle \nonumber \\
&  &   +
\tilde{U}_{21}(t;0)e^{-i(\omega -\frac{\omega_0 }{2})t} \sinh[\delta(\hat{a}-\hat{a}^{\dagger})]|1\rangle ,
\end{eqnarray}
and
\begin{eqnarray}
|\psi_g \rangle   & = &
(U_{11}(t;0)e^{-i\frac{\omega_0 t}{2}} \sinh[\delta(\hat{a}-\hat{a}^{\dagger})]+
U_{22}(t;0) e^{i\frac{\omega_0 t}{2}}\cosh[\delta(\hat{a}-\hat{a}^{\dagger})])|0 \rangle  \nonumber \\
&  &  +
\tilde{U}_{21}(t;0)e^{-i(\omega -\frac{\omega_0 }{2})t}
\cosh[\delta(\hat{a}-\hat{a}^{\dagger})]|1\rangle ,
\end{eqnarray}
and where the {\it U}'s are defined as the (vacuum) expectation values
\begin{equation}
U_{ij}(t;0)= \langle 0 | \hat{U}_{ij}(t;\hat{n}) |0 \rangle, \,\,\, i=j=1,2,
\end{equation}
and
\begin{equation}
\tilde{U}_{21}(t;0)=  -i\epsilon\langle 0 | \frac{\sin\hat{\Omega}_{\hat{n}+1}t}
{\hat{\Omega}_{\hat{n}+1}} |0 \rangle.
\end{equation}
By measuring the atom when it leaves the cavity in the state
\begin{equation}
|\Psi_{atom}  \rangle= \frac{1}{\sqrt{2}}(|g\rangle + |e\rangle) ,
\end{equation}
we end up with a wave function describing the cavity field that reads
\begin{equation}
|\Psi_{field}  \rangle= [U_{11}(t;0)e^{-i\frac{\omega_0 t}{2}} +
U_{22}(t;0)e^{i\frac{\omega_0 t}{2}}] |\delta\rangle +
\tilde{U}_{21}(t;0)e^{-i(\omega -\frac{\omega_0
}{2})t}|\delta,1\rangle ,
\end{equation}
where $|\delta,k\rangle=\hat{D}(\delta)|k\rangle$ is a displaced
number state \cite{Oli}. Therefore, we have constructed a
superpositions of displaced number states \cite{cessa}  and by
displacing the cavity field by $-\delta$, i.e. by injecting a
field that displaces the cavity field by that effective amplitude
we can generate a qubit (in reconstruction processes is a common
technique the displacement of a given wave function \cite{moya}).
\section{conclusions}
We have studied the first order contributions of the counter
rotating terms present in the interaction between a two-level atom
and a cavity field by using a technique recently introduced in
\cite{kli}. We have been able to write down the wave function in
this case, and to point out an initial state of the atom and the
field that would lead, to the generation of a quantum bit and
superpositions of displaced number states of the electromagnetic
field.

Besides the solution given here, that allows manipulation of
parameters to engineer a given state, we have looked for the
initial states to construct superpositions of displaced number
states and qubits of the electromagnetic field which are
considered highly non-classical.

It is worth to note that the qubit generated
(after displacement of the cavity field) has arbitrary amplitudes, as the coefficients
for the ground and first excited states can be varied arbitrarily. The final qubit state reads
\begin{equation}
|\Psi_{dis}  \rangle= [U_{11}(t;0)e^{-i\frac{\omega_0 t}{2}} +
U_{22}(t;0)e^{i\frac{\omega_0 t}{2}}] |0\rangle +
\tilde{U}_{21}(t;0)e^{-i(\omega -\frac{\omega_0 }{2})t}|1\rangle) .
\end{equation}
In \cite{zah} it was considered a ratio  $\lambda/\omega\approx
0.1$. Although efforts to have such  ratios (that would allow the
interaction with environments to be negligible \cite{exp}),
considering the same ratio here, and not considering the
correction we have found would indeed mislead the final result.

\acknowledgments
We would like to thank CONACYT for support.


\begin{references}


\bibitem{Knight} E.T. Jaynes and F.W. Cummings, Proc. IEEE {\bf 51}, 89 (1963); see also
B. W. Shore, and P. L. Knight,  J. Mod. Opt. {\bf 40}, 1195 (1993).
\bibitem{exp} G. Rempe, W. Schleich, M.O. Scully and H. Walther, in:
Proc. 3rd Intern. Symposium on
Foundations of Quantum Mechanics (Physical Society of Japan, Tokyo, 1989) p. 294;
M. Brune, F. Schmidt-Kaler, A. Maali, J. Dreyer, E. Hagley, J.M. Raimond, and S. Haroche,
Phys. Rev. Lett. {\bf 76}, 1800 (1996);
S. Osnaghi, P. Bertet, A. Auffeves, P. Maioli, M. Brune, J.M. Raimond, and S. Haroche,
quant-ph/0105063.
\bibitem{Dav} L. Davidovich, J.M. Raimond, M. Brune, and S. Haroche, Phys. Rev. A
{\bf 36}, 3771 (1987).
\bibitem{san} J.H. Eberly, N.B. Narozhny, and J.J. S\'anchez/Mondrag\'on, Phys. Rev. Lett.
{\bf 44}, 1323 (1980).
\bibitem{kuk} J.R. Kuklinski and J.L. Madajczyk, Phys. Rev. A {\bf 37}, 3175 (1988).
\bibitem{gea} J. Gea-Banacloche, Phys. Rev. Lett. {\bf 65}, 3385 (1990); Phys. Rev. A
{\bf 44}, 6023 (1991); V. Buzek, H. Moya-Cessa, P.L. Knight and S.J.D. Phoenix,
Phys. Rev. A {\bf 45}, 8190 (1992).
\bibitem{win1} D. M. Meekhof, C. Monroe, B. E. King, W. N. Itano, and D. J. Wineland, Phys. Rev. Lett.
{\bf 76}, 1796 (1996).
\bibitem{kis} Z. Kis, W. Vogel, and L. Davidovich, Phys. Rev. A {\bf 64}, 033401 (2001).
\bibitem{mato} R.L. de Matos Filho and W. Vogel, Phys. Rev. A {\bf 54}, 4560 (1996).
\bibitem{moyac} H. Moya-Cessa and P. Tombesi, Phys. Rev. A {\bf61}, 025401 (2000).
\bibitem{winl} D.J. Wineland, C. Monroe, W.M. Itano, D. Leibfried, B.E. King, and
and D.M. Meekhof, J. Res. Nat. Inst. Stan. {\bf 103}, 259 (1998).
\bibitem{eke} A. Ekert and C. Macchiavello, Acta Phys. Polonica A {\bf 93}, 63 (1998).
\bibitem{moy3} H. Moya-Cessa, D. Jonathan, and P.L. Knight,
e-print quant-ph/0110167.
\bibitem{zah1} K. Zaheer and M.S. Zubairy, Phys. Rev. A,  {\bf 37}, 1628 (1988).
\bibitem{zah} K. Zaheer and M.S. Zubairy, Opt. Commun. {\bf 73}, 325 (1989).
\bibitem{cessa} H. Moya-Cessa, J. of Mod.
Optics {\bf 42}, 1741 (1995).
\bibitem{kli} A. Klimov and L.L. S\'anchez-Soto, Phys. Rev. A, {\bf 61}, 063802 (2000).
\bibitem{Oli} F.A.M. de Oliveira, M.S. Kim, P.L. Knight, and V. Buzek,
Phys. Rev. A {\bf 41}, 2645 (1990).
\bibitem{moya} L.G. Lutterbach and L. Davidovich, Phys. Rev. Lett. {\bf 78}, 2547 (1997);
H.Moya-Cessa, J.A. Roversi, S.M. Dutra, and A.Vidiella-Barranco, Phys. Rev. A {\bf 60}, 4029 (1999).
\end{references}
\end{document}